# Spin-pump-induced spin transport in p-type Si at room temperature

Eiji Shikoh,[1,*,†] Kazuya Ando,[2] Kazuki Kubo,[1] Eiji Saitoh,[2,3,4] Teruya Shinjo,[1] and Masashi Shiraishi[1,5,*,‡]

[1]*Graduate School of Engineering Science, Osaka University, Toyonaka 560-8531, Japan*

[2]*Institute for Materials Research, Tohoku University, Sendai 980-8577, Japan*

[3]*CREST, Japan Science and Technology Agency, Sanbancho, Tokyo 102-0075, Japan*

[4]*The Advanced Science Research Center, Japan Atomic Energy Agency, Tokai 319-1195, Japan*

[5]*PRESTO, Japan Science and Technology Agency, Sanbancho, Tokyo 102-0075, Japan*

A spin battery concept is applied for the dynamical generation of pure spin current and spin transport in p-type silicon (p-Si). Ferromagnetic resonance and effective s-d coupling in $Ni_{80}Fe_{20}$ results in spin accumulation at the $Ni_{80}Fe_{20}$/p-Si interface, inducing spin injection and the generation of spin current in the p-Si. The pure spin current is converted to a charge current by the inverse spin Hall effect of Pd evaporated onto the p-Si. This approach demonstrates the generation and transport of pure spin current in p-Si at room temperature.

Pure spin current is one of the most attractive phenomena in spintronics. To date, three approaches have been used to generate pure spin currents: electrical, thermal, and dynamical. The electrical method is the most commonly used for estimating spin coherence in inorganic and carbon-based materials [1,2]. The thermal method, on the other hand, is used to generate spin current via spin-wave induction in YIG with long-range spin coherence [3]. However, it was proposed theoretically that a spin battery - a device that yields pure spin current via ferromagnetic resonance (FMR) - can dynamically generate pure spin current for the estimation of spin coherence [4], but this has not yet been achieved. Currently, the use of the dynamical method is limited to the modulation of the Gilbert damping constant [5,6] and spin injection in materials with strong spin-orbit coupling [3,7-12]. Nevertheless, the dynamical method enables the circumvention of the conductance mismatch problem [13], and is therefore effective for realizing the transport of a pure spin current in materials with low conductivity, such as silicon (Si).

Si is becoming an increasingly desired material for spintronics [12,14-25] because it is a pivotal material in today's semiconductor electronics industry. Meanwhile, the understanding of spin transport in Si, particularly p-type Si (p-Si), remains incomplete. Whereas the spin-orbit interaction (SOI) in p-Si cannot be ignored, the spin split-off energy in Si (0.04 eV) [26] is much smaller than in GaAs (0.34 eV) [27], which suggests that spin transport in p-Si may be experimentally realized. In this study, we used the dynamical method, based on the spin battery, to generate pure spin current in p-Si. We achieved efficient spin transport in p-Si (see Fig. 1(a)). The pumped spins propagate in the p-Si as a pure spin current, and are absorbed in a Pd wire onto the p-Si, where the spin current $\mathbf{J}_s$ is converted to a charge current $\mathbf{J}_c$ by an inverse spin

Hall effect (ISHE) [7], as $\mathbf{J}_c = D_{ISHE}\,\mathbf{J}_s \times \boldsymbol{\sigma}$. Here, $D_{ISHE}$ is a coefficient representing the ISHE efficiency in the material, $\mathbf{J}_s$ lies along the normal direction to the Pd wire plane (to the p-Si substrate plane), and the dc component of $\boldsymbol{\sigma}$ is parallel to the magnetization-precession axis in the $Ni_{80}Fe_{20}$ (Py) film, as shown in Fig. 1(a). Spin pumping has been widely recognized as a potential tool for injecting spins [3,5,7-12], and we shed light upon a novel aspect of spin pumping phenomena. We report that spin pumping can generate a pure spin current in condensed matter. Specifically, spin-pumping-induced spin transport in condensed matter was demonstrated at room temperature (RT).

Figures 1(b) and 2(c) show schematic illustrations of a Py/p-Si/Pd film sample consisting of a 25-nm-thick Py layer, a 100-nm-thick p-Si layer with a boron (B) doping concentration $N_a$ of $1.0 \times 10^{19}$ $cm^{-3}$, and a 5-nm-thick paramagnetic Pd wire. This sample was used in the main experiments involving spin injection and transport in the p-Si. The definitions of the applied external magnetic field angle $\theta_H$ and the Py magnetization-orientation angle $\theta_M$ are shown in Fig. 2(c). Just before the metal films were formed, the surface of the p-Si substrate was etched by hydrofluoric acid to remove the naturally-oxidized Si. The Pd wire and the Py film were formed by electron beam lithography and electron beam deposition. The surface area of the Py film was $900 \times 150$ $\mu m^2$, and the width of the Pd wire was 1 μm. Two contacts were attached to the Pd wire with Ag paste to enable inverse spin-Hall effect (ISHE) voltage measurements. The distance $L$ between the Py film and the Pd wire was estimated to be 490 nm using atomic force microscopy (AFM). The spin transport in the p-Si at RT was detected using the ISHE in the Pd wire [7,8]. Figure 2(a) shows a schematic illustration of a 25-nm-thick Py film sample, which was used in experiments to investigate the magnetic properties of a Py spin injector. The Py film was formed by electron beam lithography and electron beam deposition on a thermally oxidized Si substrate. Each sample was placed near the center of a $TE_{011}$ microwave cavity of an electron resonance system with a frequency of $f = 9.44$ GHz. All measurements were performed at RT.

Figure 2(b) shows the FMR spectra $dI(H)/dH$ of a Py/$SiO_2$ sample (black line) and a Py/p-Si/Pd sample (red line) at $\theta_H = 0$ with a microwave excitation of 100 mW. The ferromagnetic resonance fields $H_{FMR}$ for the Py/$SiO_2$ and Py/p-Si/Pd samples were 130 mT and 131 mT, respectively. The spectral width $W$ in the FMR signals was 4.42 mT for the Py/p-Si/Pd sample, which was greater than that of the Py/$SiO_2$ sample, 3.49 mT. Figure 2(d) shows the dc electromotive force signal $V$ for the Py/p-Si/Pd sample at $\theta_H = 0$ with a microwave excitation of 100 mW. The open circles and blue line in Fig. 2(d) represent the experimental data and theoretical fitting line, respectively. The theoretical fitting function [7] is described as

$$V(H) = I_{ISHE}\frac{\Gamma^2}{(H-H_{FMR})^2+\Gamma^2} + I_{Asym}\frac{-2\Gamma(H-H_{FMR})}{(H-H_{FMR})^2+\Gamma^2}, \qquad (1)$$

where $\Gamma$ denotes the damping constant (4.3 mT in this study) and $H_{FMR}$ is the ferromagnetic resonance field (131 mT). The first and second terms in eq. (1) correspond to the contributions from the ISHE and the asymmetry term against $H$ (ex. anomalous-Hall effect), respectively. $I_{ISHE}$ and $I_{Asym}$ correspond to the coefficients of the first and second terms in eq. (1). The ratio of the $I_{ISHE}$ to $I_{Asym}$ in eq. (1) was calculated to be 6.4 according to the fitting shown in Fig. 2(d), corroborating the assumption that the observed signal can be attributed to the ISHE in the Pd wire [7,10]. The microwave power dependence of $V$ for the Py/p-Si/Pd sample was investigated at $\theta_H = 0$ (See Supplemental Material). $I_{ISHE}$ was almost proportional to the microwave power, which is plausible because the density of the generated spin current in the p-Si proportionally increases with increasing applied microwave power [8,9]. The microwave power dependence of the electromotive force provides additional evidence that the observed voltage signals can be ascribed to the ISHE in the Pd wire.

Figures 3(a) and 3(b) show the $\theta_H$ dependence of the FMR signals $dI(H)/dH$ and of the dc electromotive force in the Pd wire of the Py/p-Si/Pd sample under a microwave excitation of 100 mW, respectively. The open circles and solid lines in Fig. 3(b) are the experimental data and theoretical fitting lines based on eq. (1), respectively. A more detailed $\theta_H$ dependence of the electromotive force normalized by the value at $\theta_H = 0$ is shown in Fig. 3(c). A clear angular dependence of the electromotive force was observed; that is, the sign of the electromotive force was inverted by the magnetization reversal of the Py film, in good accordance with the ISHE theory [7]. The static equilibrium condition yields an expression [11] that relates $\theta_H$ and $\theta_M$ as follows, $(H_{FMR}/4\pi M_s)\sin(\theta_H - \theta_M) = \sin\theta_M \cos\theta_M$, where $H$ and $M_s$ are the strength of the external magnetic field and the saturation magnetization of the Py layer, respectively (See Supplemental Material). The electromotive force was replotted as a function of $\theta_M$, as shown in Fig. 3(d). In Fig. 3(d), the solid circles show experimental data normalized by the value at $\theta_M = 0$, and the dashed line is a fitting curve proportional to $\cos\theta_M$. Figure 3(d) suggests that the electromotive force is proportional to $\cos\theta_M$, demonstrating again that the relationship $\mathbf{J}_c = D_{ISHE}\,\mathbf{J}_s \times \boldsymbol{\sigma}$ is relevant to ISHE in this present system [11]. This finding indicates that the ISHE in the Pd wire can be attributed to propagating spins in the p-Si. Although the SOI in the p-Si is not small enough, it should be emphasized that the observed ISHE signal cannot be ascribed to the ISHE in the p-Si [12]. No signal of the ISHE in the p-Si was observed in our sample geometry, probably because our sample does not utilize the Si-on-insulator structure, so the ISHE signal in the p-Si is too weak.

We prepared a sample with both Pd and Cu wires (Py/p-Si/Pd&Cu) for a control experiment. Figure 4(a) shows a schematic illustration of a Py/p-Si/Pd&Cu sample with the $\theta_H$ direction defined. The gap length between the Py film and the Pd or Cu wire; that is, the spin transport length *L,* was estimated to be 690 nm using AFM. Figure 4(b) shows the $\theta_H$ dependence of the dc electromotive force in the Pd wire (purple circles), and in the Cu wire (green circles) of the Py/p-Si/Pd&Cu sample under a microwave excitation of 200 mW. It is noteworthy that the electromotive force in the Pd wire was observed as expected, but in the same sample no signals were observed in the Cu wire, which has a smaller SOI than Pd. This strongly indicates that the output voltages in Figs. 2(d) and 3(b) can be attributed to the ISHE in the Pd. Moreover, the carrier doping concentration in the p-Si and the temperature dependences of the electromotive forces in the Py/p-Si/Pd samples were investigated, and these also showed decisive evidence of spin transport in the p-Si (See Supplemental Material).

We can estimate the spin diffusion length $\lambda$ in the p-Si from the experimental data shown in Figs. 2(b) and 2(d). Since a Schottky barrier reduces spin injection and detection efficiency, $\lambda$ was theoretically estimated using a simple model without taking the *s-d* coupling at the Py/p-Si interface into account [8]. The dynamics of the magnetization $M(t)$ in the Py under an effective magnetic field $H_{\mathrm{eff}}$ is described by the Landau-Lifshitz-Gilbert equation

$$\frac{dM(t)}{dt} = -\gamma M(t) \times H_{eff} + \frac{\alpha}{M_s} M(t) \times \frac{dM(t)}{dt}, \qquad (2)$$

where $\gamma$, $\alpha$, and $M_S$ are the gyromagnetic ratio, damping constant due to another definition [5], and the saturation magnetization of the Py, respectively. The dynamical magnetization process induces spin pumping from the Py to the p-Si and generated a spin current $j_s$, as

$$j_s = \frac{\omega}{2\pi}\int_0^{2\pi/\omega} \frac{\hbar}{4\pi} g_r^{\uparrow\downarrow} \frac{1}{M_s^2}[M(t)\times\frac{dM(t)}{dt}]_z dt. \tag{3}$$

Here, $g_r^{\uparrow\downarrow}$ and $\hbar$ are the real part of the mixing conductance [6] and the Dirac constant, respectively. Note that $g_r^{\uparrow\downarrow}$ in eq. (3) is given by

$$g_r^{\uparrow\downarrow} = \frac{2\sqrt{3}\pi M_s \gamma d_F}{g\mu_B\omega}(W_{Py/Si} - W_{Py}), \tag{4}$$

where $g$, $\mu_B$, and $d_F$ are the g-factor, the Bohr magneton, and thickness of the Py layer, respectively. $d_F$, $W_{Py/Si}$, and $W_{Py}$ were 25 nm, 4.42 mT, and 3.49 mT, respectively. From the equations, the spin current density at the Py/p-Si interface is obtained as

$$j_s = \frac{g_r^{\uparrow\downarrow}\gamma^2 h^2 \hbar[4\pi M_s\gamma + \sqrt{(4\pi M_s)^2\gamma^2 + 4\omega^2}\,]}{8\pi\alpha^2[(4\pi M_s)^2\gamma^2 + 4\omega^2]}, \tag{5}$$

where $h$ is the microwave magnetic field, set to 0.11 mT at a microwave power of 100 mW. The total $g_r^{\uparrow\downarrow}$ via the two interfaces (Py/p-Si and p-Si/Pd) was calculated to be $1.74\times 10^{19}$ m$^{-2}$, and thus $j_s$ in the Pd was calculated to be $1.13\times 10^{-8}$ Jm$^{-2}$. Given our device geometry, we assume that half of the generated $j_s$ contributes to the electromotive force in the Pd wire. The conductivity of the p-Si with a doping concentration of $1\times 10^{19}$ cm$^{-3}$ ($1.1\times 10^4$ $(\Omega\mathrm{m})^{-1}$) is two orders of magnitude lower than that of the Pd ($1.97\times 10^6$ $(\Omega\mathrm{m})^{-1}$) [8], and thus the electromotive force, taking the spin relaxation in the Pd wire into account, is approximately written as

$$V_{ISHE} = \frac{w\theta_{SHE}\lambda_{Pd}\tanh(d_{Pd}/2\lambda_{Pd})}{d_{Pd}\sigma_{Pd}}(\frac{2e}{\hbar})j_s\ , \tag{6}$$

where $w$, $\lambda_{Pd}$, $d_{Pd}$, and $\sigma_{Pd}$ are the length of the Pd wire facing the Py (900 μm), the spin diffusion length (9 nm [28]), the thickness (5 nm), and the conductivity of the Pd, respectively. The spin-Hall angle $\theta_{SHE}$ from Py to Pd has been reported to be 0.01 [8], which allows us to theoretically estimate the electromotive force in the Pd wire as $3.84\times10^{-5}$ V if half of the spin current, generated by the spin pumping, contributes to the electromotive force in the Pd wire, as described above. In contrast, the experimentally observed electromotive force was $1.4\times10^{-6}$ V and the discrepancy can be ascribed to dissipation of the spin current during spin transport in the p-Si, which can be described as an exponentially damping dependence on the spin transport length. By ignoring the latter effect, $\lambda$ in the p-Si at RT was estimated to be 148 nm because the $L$ was measured to be 490 nm. It should be emphasized that $\lambda$ can be readily enhanced by improving the device fabrication process and removing the natural oxide layer, which we will attempt in the near future. The charge diffusion constant $D$ of the p-Si was calculated to be 1.8 $cm^2/s$ from Einstein's relationship, so the spin coherent time $\tau$ was estimated to be 122 ps from the relationship $\lambda=\sqrt{D\tau}$ [19], under the assumption that the charge and spin diffusion constants are the same.

$\lambda$ in p-Si with a conductivity of $9.1\times10^{3}$ $(\Omega m)^{-1}$ has previously been reported to be 310 nm at RT by observing 3-terminal spin accumulation, an electrical spin injection and spin accumulation method [19]. The $\lambda$ values in the previous and present studies are relatively close, however it is noteworthy that the $\lambda$ values in the previous study [19] were not estimated by directly observing spin transport. Therefore, our study provides the first demonstration of spin transport, and simultaneously represents the first estimation of $\lambda$ in p-Si at RT by directly observing spin transport phenomena. It is well known that the spin accumulation voltages in the 3-terminal method are easily enhanced by the existence of carrier traps at the interface [29]. Therefore, discussion of the physical meaning of the non-local 3-terminal method is underway, and we have reported possible contributions from spurious effects [25]. Thus, the estimation of $\lambda$ using spin pumping after achieving spin transport is quite significant, and has been eagerly awaited. We emphasize that our estimation provides both $\lambda$ and $\tau$, and that better spin coherence in p-Si at RT is expected by improving the spin-injector/p-Si interface. $\tau$ in our study seems slightly longer than expected from theory [24], judging from momentum relaxation time. This remains an open question, and understanding the spin transfer mechanism from Py to the Pd requires additional investigation, which is unfortunately beyond the scope of this manuscript. The discrepancy will be investigated in detail in the near future.

In summary, we experimentally demonstrated spin-pumping-induced spin current generation and spin transport in highly-doped p-Si at RT using the spin battery concept. The spin diffusion length in the p-Si was estimated to be ca. 150 nm at RT.

The authors (E. Sh. and M. S.) thank Yoshiya Honda (Osaka University) for his assistance in sample preparation. This research was partly supported by a Grant-in-Aid for Scientific Research from the MEXT, Japan and by the Global COE program "Core Research and Engineering of Advanced Materials Interdisciplinary Education Center for Materials Research".

*These two authors contributed equally to this work.

†shikoh@ee.es.osaka-u.ac.jp

‡shiraishi@ee.es.osaka-u.ac.jp

FIG. 1 (color online) (a) A schematic illustration of spin-pumping-induced spin current generation in p-Si (substrate) and the conversion of spin current to charge current in a paramagnetic Pd wire due to the inverse spin-Hall effect (ISHE) [7]. In the FMR condition of the $Ni_{80}Fe_{20}$ (Py) film, the dynamical exchange interaction drives the spin pumping, spin propagation in the p-Si, and absorption of pure spin currents into the Pd. (b) A schematic illustration of the sample, consisting of a Py film and a Pd wire on a p-Si substrate. $L$ is the gap length between the Py film and the Pd wire. (c) An optical microscope image of a part of the sample surface.

FIG. 2 (color online) (a) A schematic illustration of an $Ni_{80}Fe_{20}$(Py) film sample on an $SiO_2$ substrate. (b) The external magnetic field $H$ dependence of the FMR signals $dI(H)/dH$ measured for the Py/$SiO_2$ sample (a black line) and for the Py/p-Si/Pd sample (a red line) shown in Fig. 2(c) when $H$ was applied parallel to the film plane. Here, $I$ denotes the microwave absorption intensity, and the definition of the spectral width $W$ is shown in the inset of the figure. (c) A schematic illustration of the sample consisting of a Py film and a Pd wire on a p-Si substrate, with the definition of the magnetic field angle $\theta_H$ and the Py magnetization-orientation angle $\theta_M$. (d) $H$ dependence of the electromotive force $V$ measured for the Py/p-Si/Pd sample at $\theta_H = 0$. The open circles are the experimental data, and the blue solid line shows the fitting results based on eq. (1).

FIG. 3 (color online) (a) $\theta_H$ dependence of the FMR signals $dI(H)/dH$ measured for the $Ni_{80}Fe_{20}$(Py)/p-Si/Pd sample. $H_{FMR}$ was 131 mT at $\theta_H = 0$, 1142 mT at $\theta_H = 90°$, and 130 mT at $\theta_H$ = 180°. (b) $\theta_H$ dependence of the electromotive force $V(\theta_H)$ measured for the Py/p-Si/Pd sample. The open circles are the experimental data and the solid lines show fitting results based on eq. (1). (c) $\theta_H$ dependence of the electromotive force $V$ normalized by the $V$ value at $\theta_H = 0$. (d) $\theta_M$ dependence of the electromotive force $V$ normalized by the $V$ value at $\theta_M = 0$. The dashed line is a fitting curve proportional to cos $\theta_M$.

FIG. 4 (color online) (a) A schematic illustration of the $Ni_{80}Fe_{20}$(Py)/p-Si/Pd&Cu sample, and the definition of the magnetic field angle $\theta_H$. $L$ is the gap length between the Py film and the Pd or Cu wire. (b) $\theta_H$ dependence of the electromotive force $V(\theta_H)$ in the Pd wire (purple circles), and that in the Cu wire (green circles) of the Py/p-Si/Pd&Cu sample.

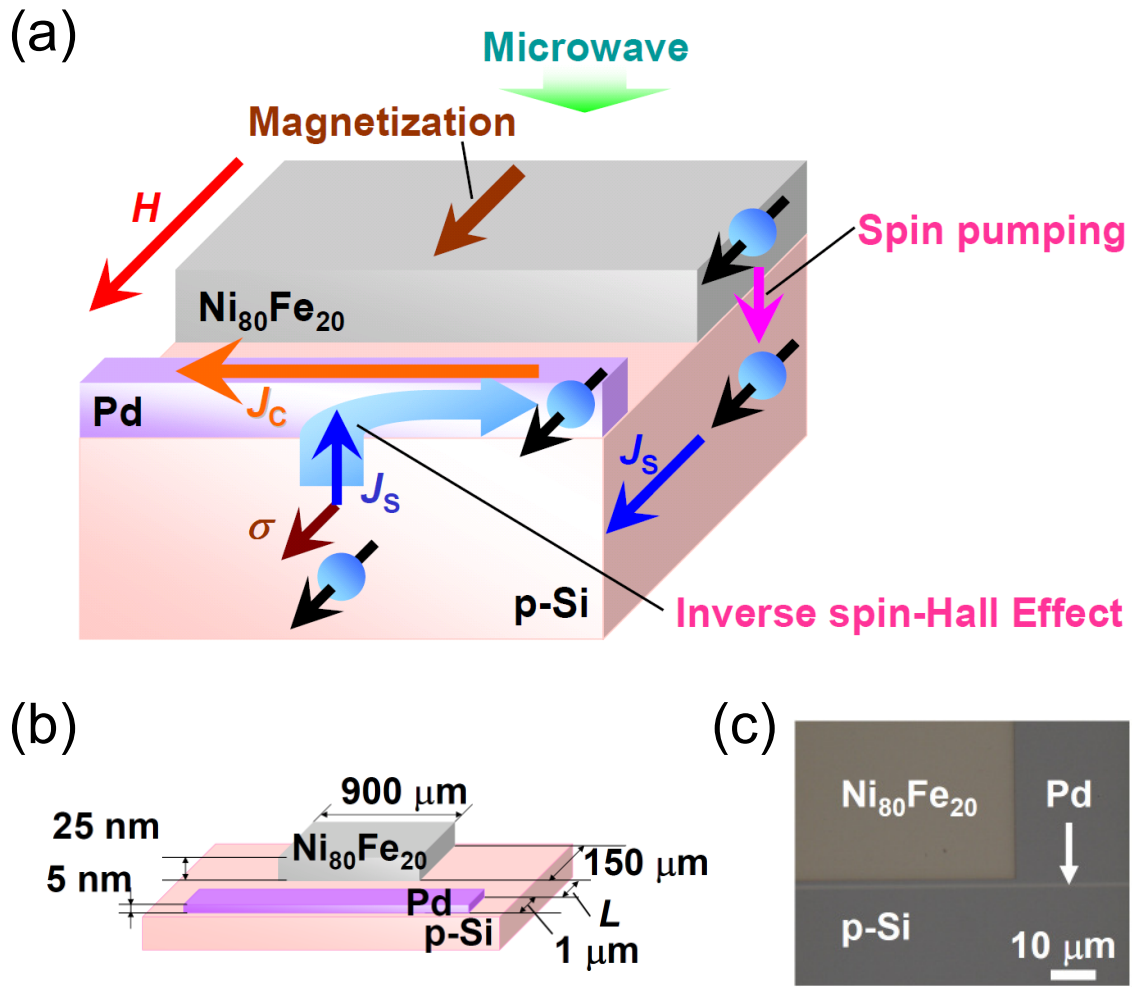

E. Shikoh, *et al.*, FIG 1.

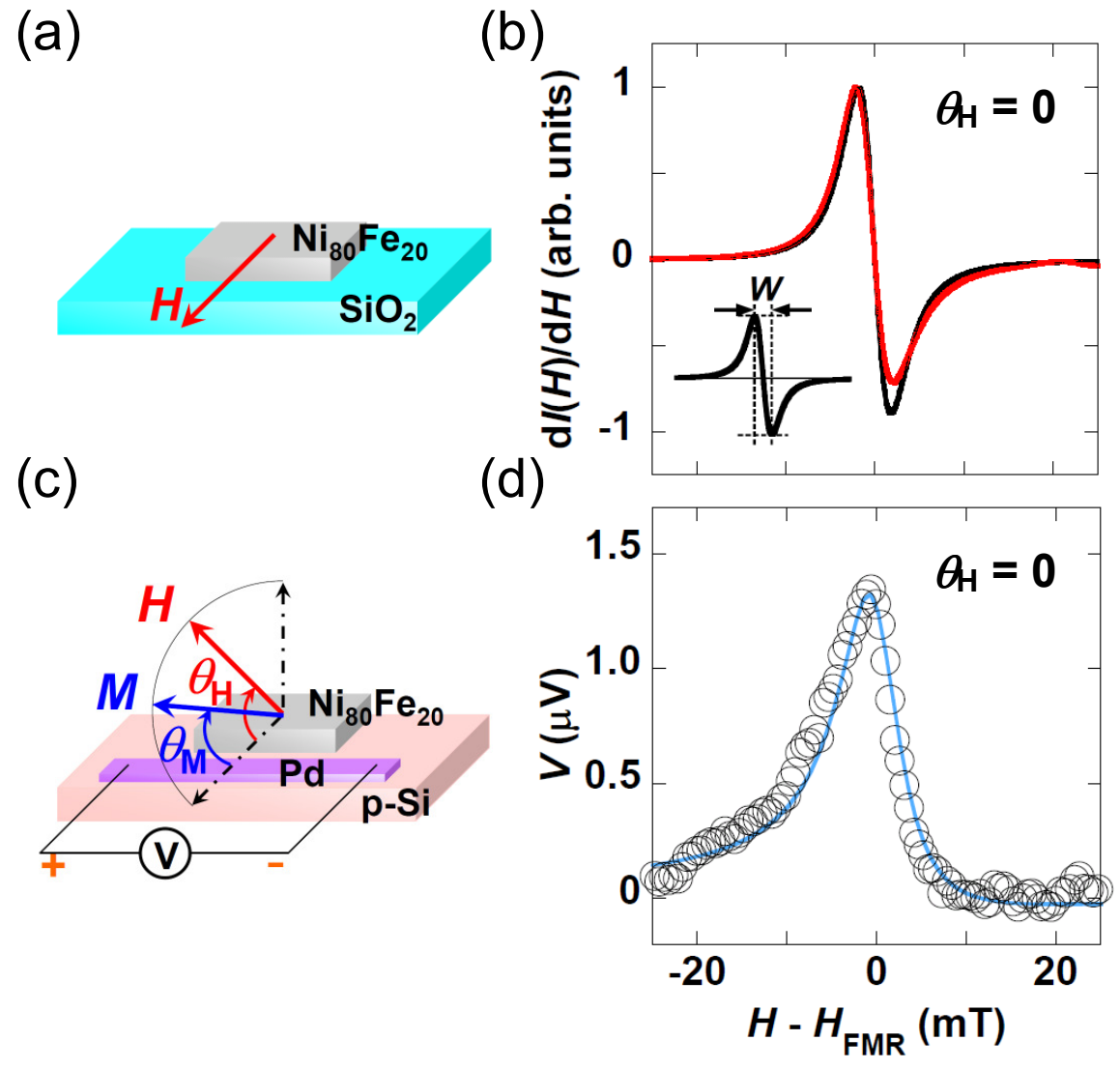

E. Shikoh, *et al.*, FIG. 2.

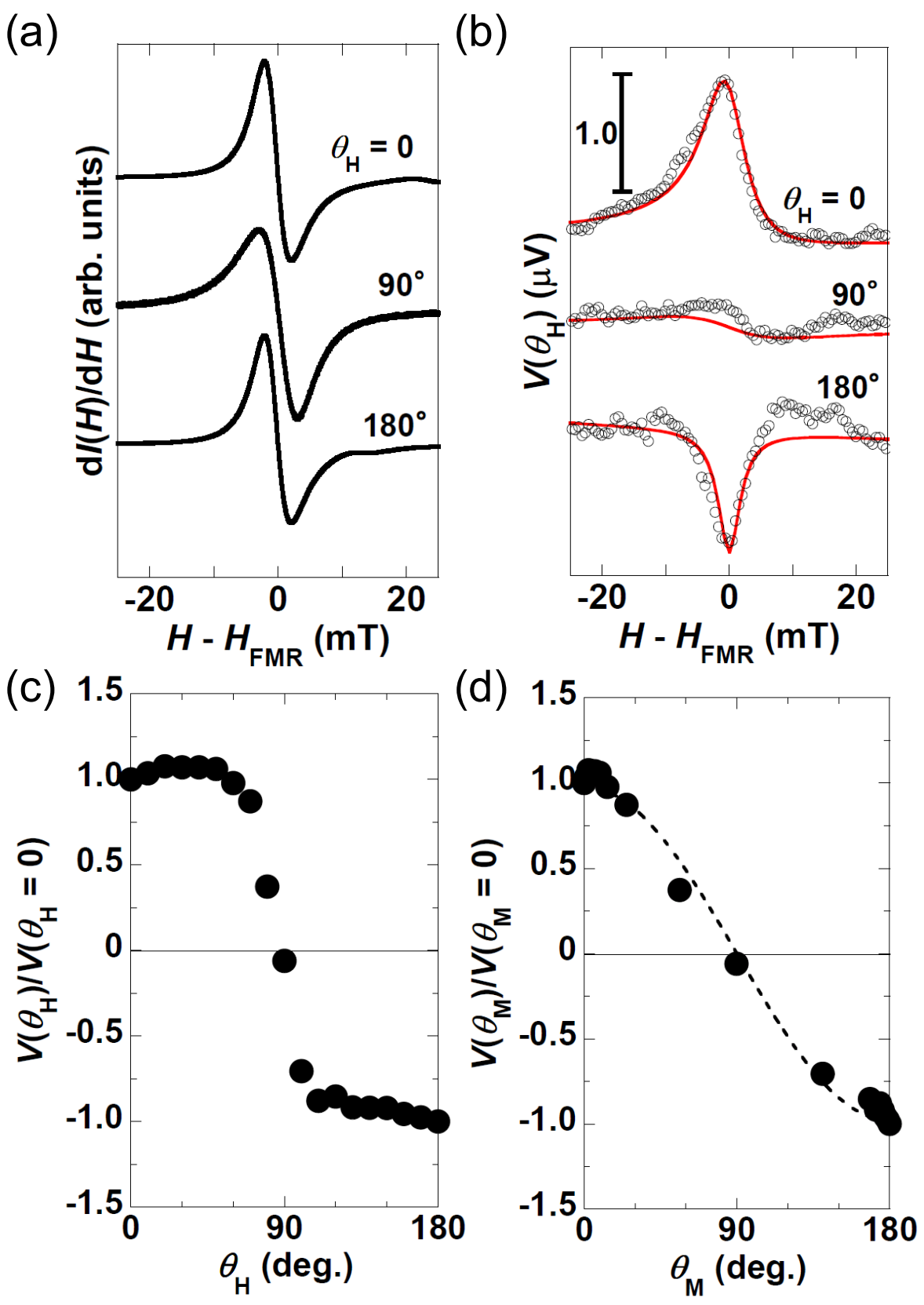

E. Shikoh, *et al.*, FIG. 3.

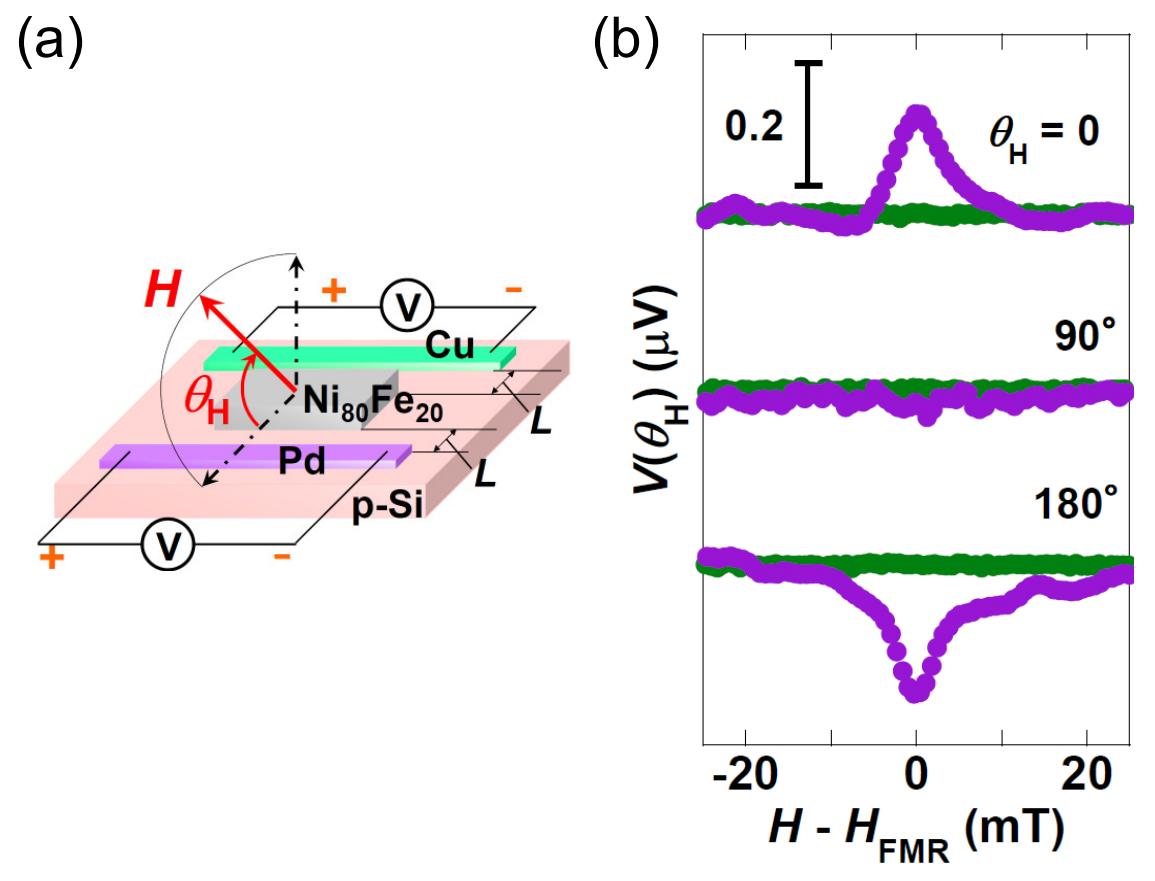

E. Shikoh, *et al.*, FIG. 4.

Supplemental Material

# Spin-pumping-induced spin transport in p-type Si at room temperature

Eiji Shikoh, Kazuya Ando, Kazuki Kubo, Eiji Saitoh, Teruya Shinjo, and Masashi Shiraishi

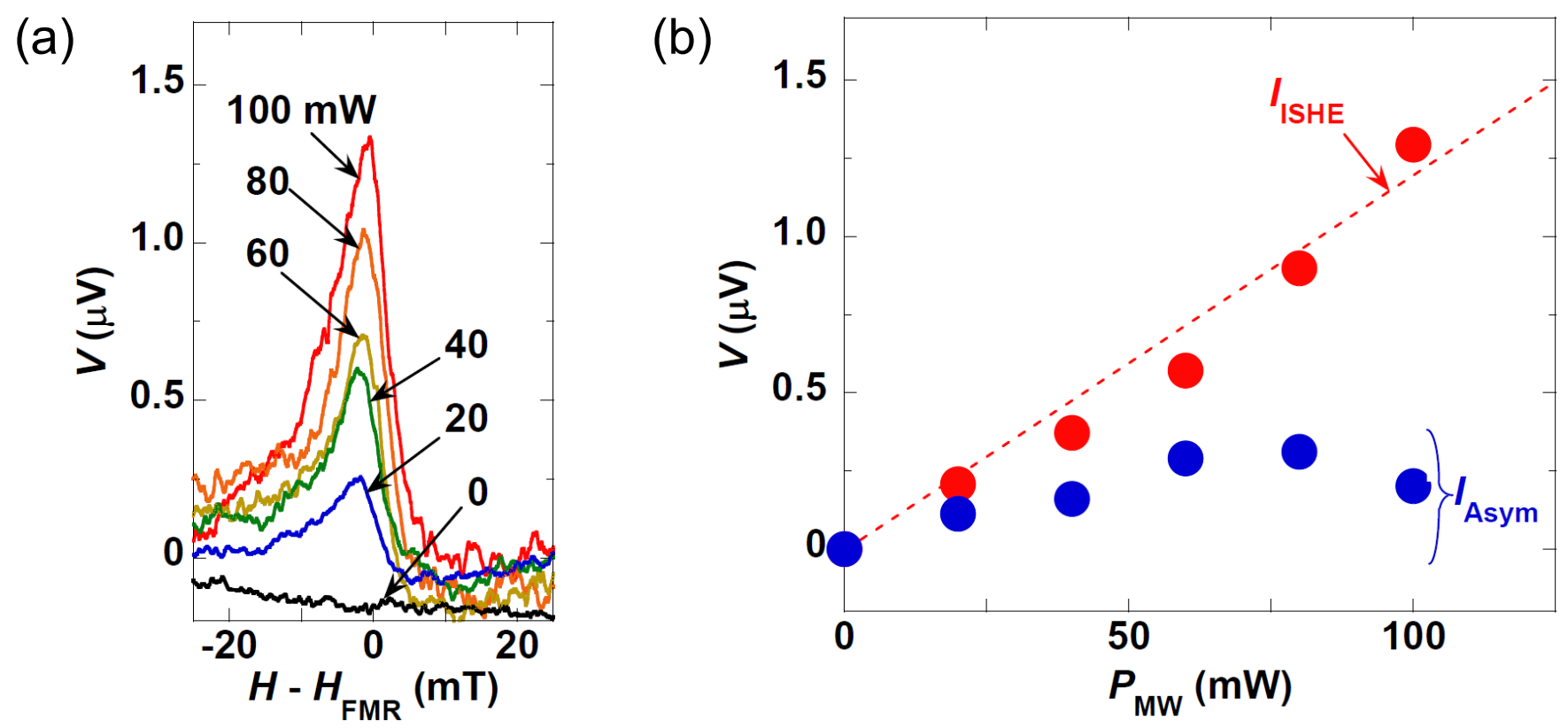

FIG. S1. (a) External magnetic field $H$ dependence of the electromotive force $V$ measured for the $Ni_{80}Fe_{20}$(Py)/p-Si/Pd sample at $\theta_H = 0$, under different microwave excitation powers. (b) Microwave power ($P_{MW}$) dependence of $I_{ISHE}$ and $I_{Asym}$, measured for the Py/p-Si/Pd sample, and analyzed using eq. (1) in the main text. The dashed line shows the linear fit of the data for $I_{ISHE}$.

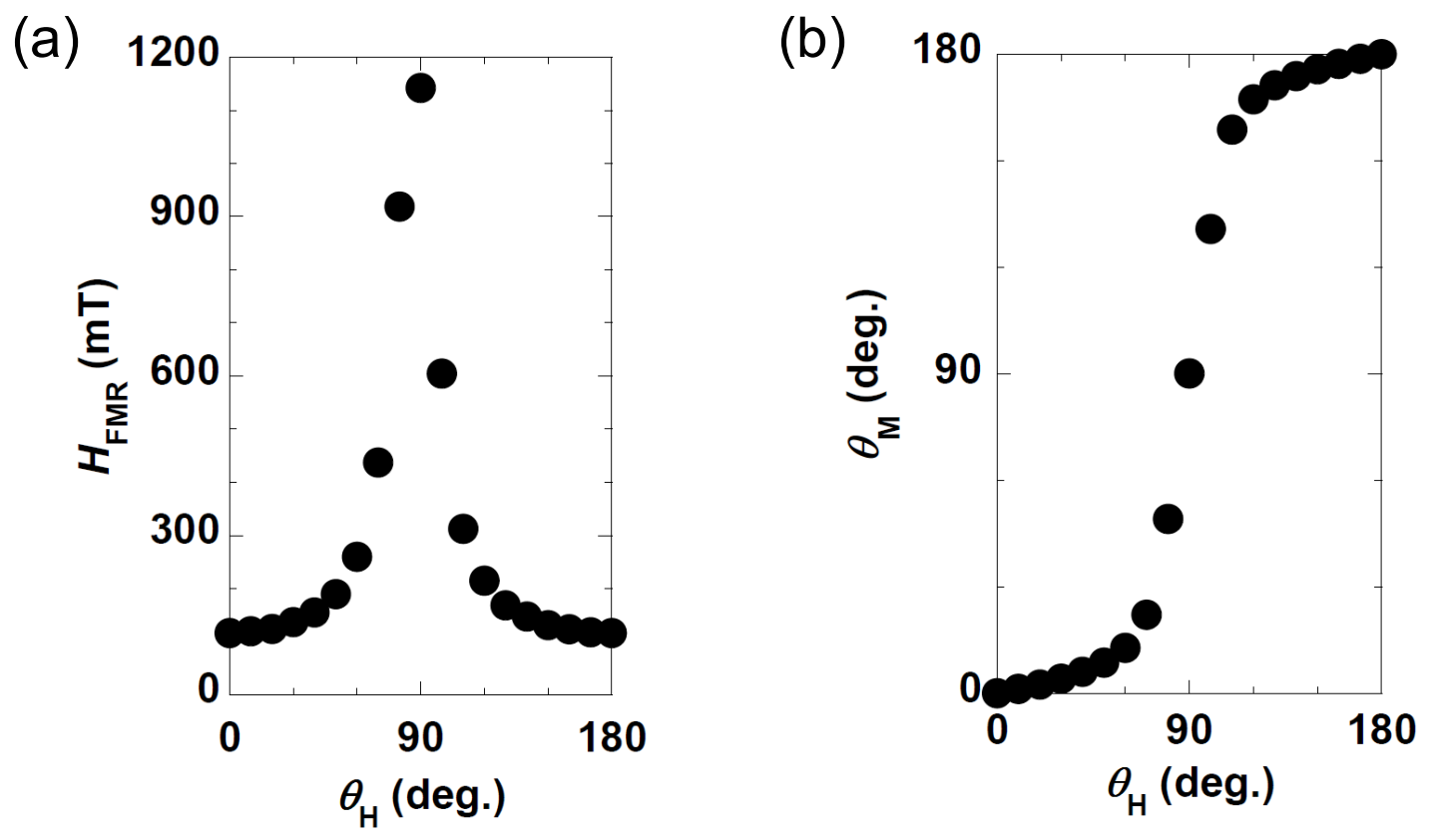

FIG. S2. (a) $\theta_H$ dependence of the ferromagnetic resonance field $H_{FMR}$. (b) Relationship between $\theta_H$ and $\theta_M$.

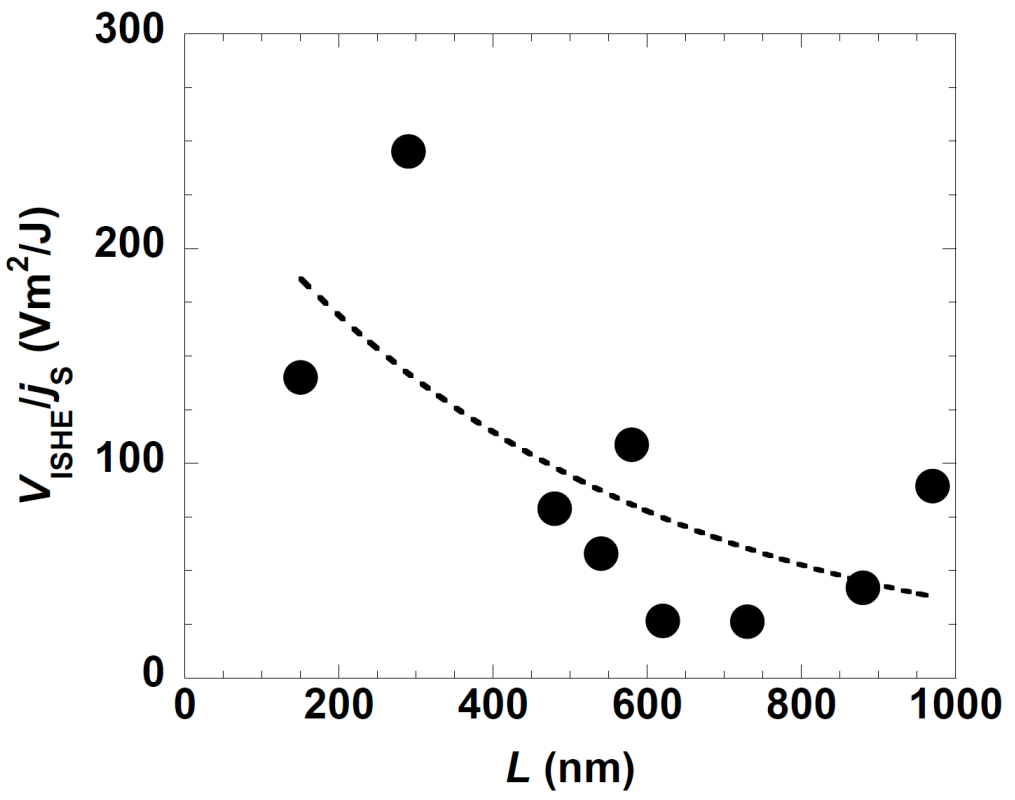

FIG. S3. Gap length ($L$) dependence of the electromotive force in the Pd wire of the $Ni_{80}Fe_{20}$/p-Si/Pd samples. Solid circles are output voltage signals normalized by the generated spin current $j_s$. The dashed line is a fitting curve using an exponential function.

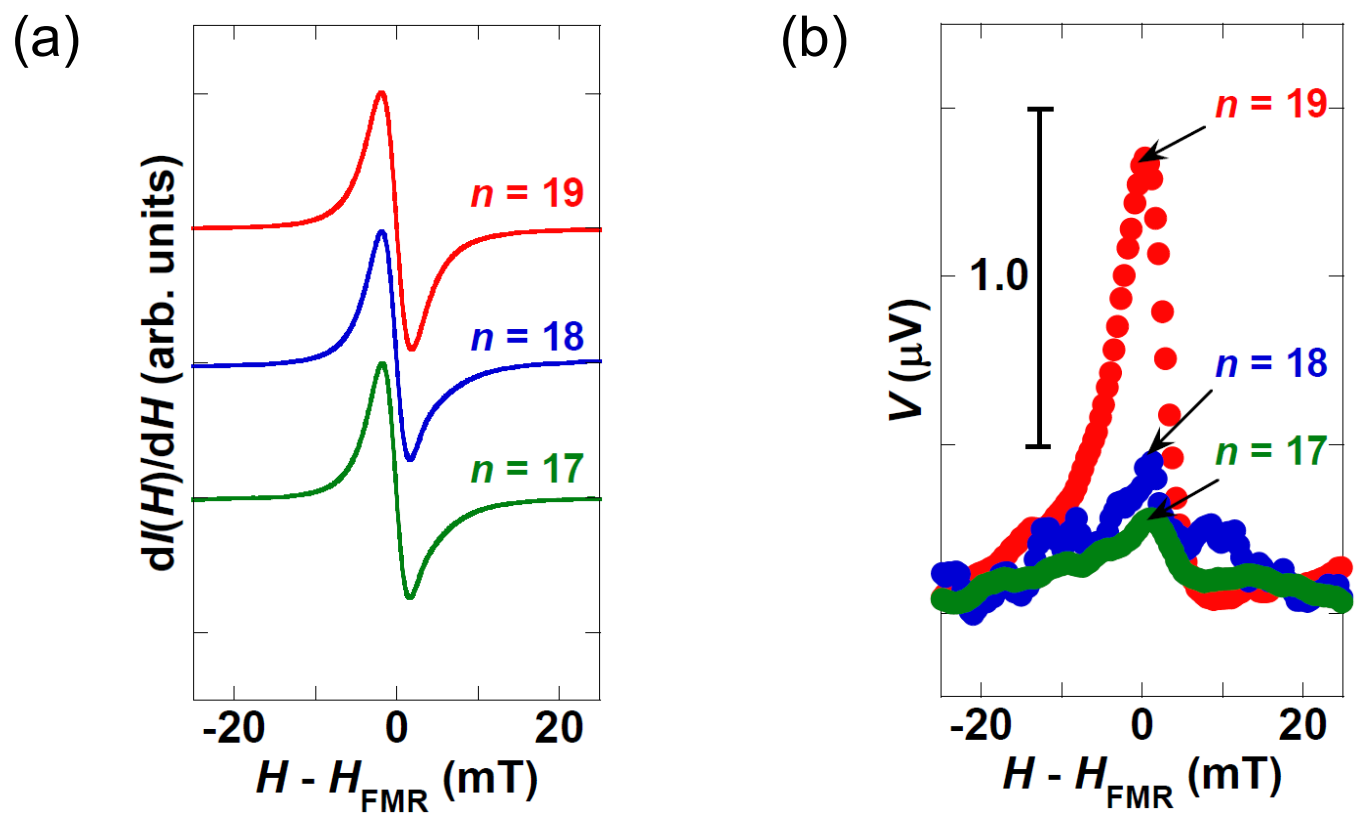

FIG. S4. (a) FMR signals with a carrier doping concentration in p-Si of $1.0 \times 10^{n}$ $cm^{-3}$ (n = 17, 18 and 19) in $Ni_{80}Fe_{20}$(Py)/p-Si/Pd samples with $L$ = 490 nm, $\theta_H$ = 0, and a microwave power of 100 mW. The red, blue, and green lines correspond to the FMR spectrum with $n$ values of 19, 18, and 17, respectively. (b) Electromotive force $V$ with a carrier doping concentration in the p-Si of $1.0 \times 10^{n}$ $cm^{-3}$ (n = 17, 18 and 19), $\theta_H$ = 0, and a microwave power of 100 mW. The red, blue, and green closed circles correspond to the electromotive force with $n$ values of 19, 18, and 17, respectively.

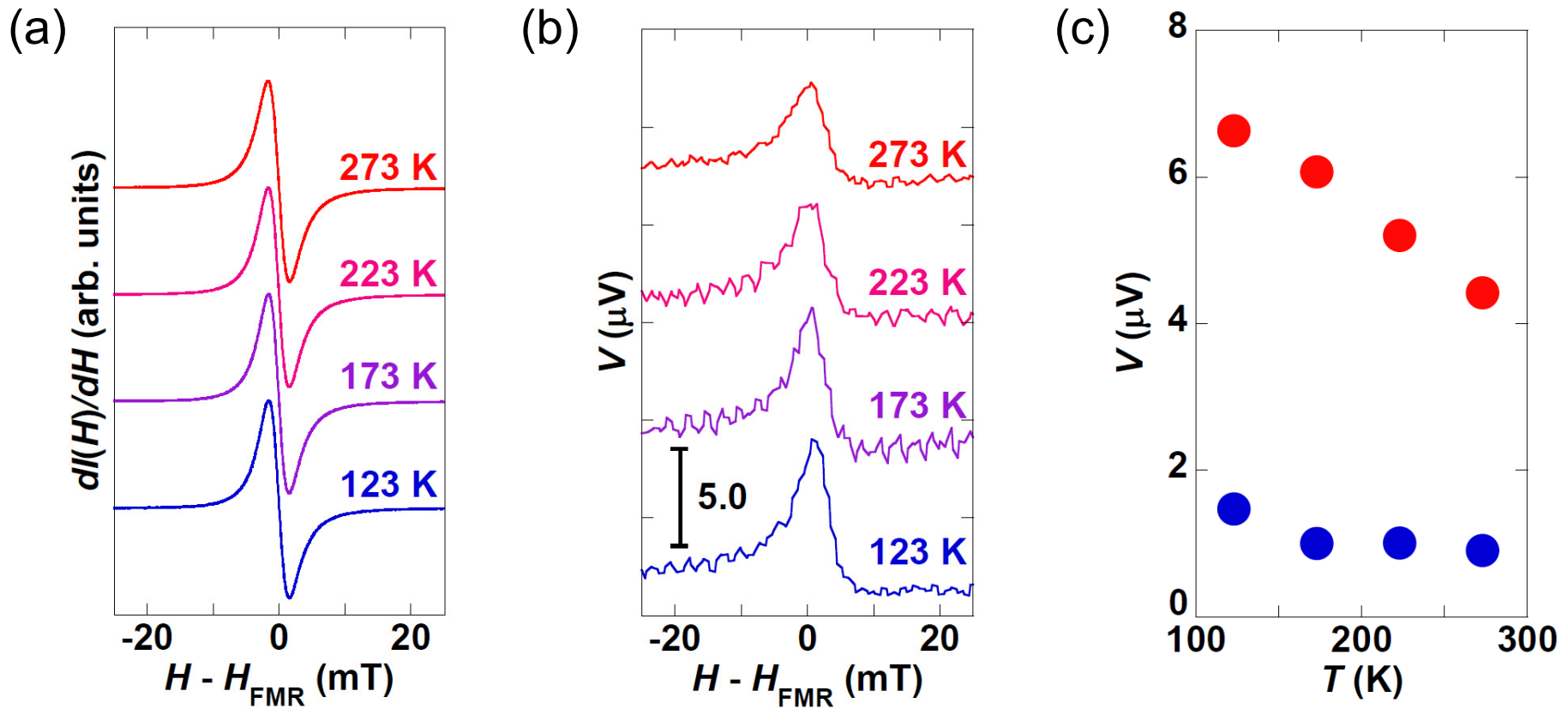

FIG. S5. Temperature dependences of (a) FMR signals and (b) electromotive forces in $Ni_{80}Fe_{20}$(Py)/p-Si/Pd samples with $L$ = 580 nm, $\theta_H$ = 0, and a microwave power of 200 mW. (c) Temperature dependences of $I_{ISHE}$ (red closed circles) and $I_{Asym}$ (blue closed circles), measured for the Py/p-Si/Pd sample with $L$ = 580 nm, and analyzed using eq. (1) in the main text.